\documentclass[final,3p,times]{elsarticle} 
\usepackage{amssymb,amsmath,graphicx,booktabs,hyperref}

\DeclareMathOperator{\sech}{sech}
\DeclareMathOperator{\cn}{cn}
\DeclareMathOperator{\round}{round}

\journal{Mathematics and Computers in Simulation}

\begin{document}

\begin{frontmatter}

\title{Hidden solitons in the Zabusky--Kruskal experiment:\\ Analysis using the periodic, inverse scattering transform}

\author{Ivan C. Christov}
\ead{christov@alum.mit.edu}
\ead[url]{http://alum.mit.edu/www/christov}
\address{Department of Engineering Sciences and Applied Mathematics, Northwestern University, Evanston, IL 60208-3125, USA}

\begin{abstract}
Recent numerical work on the Zabusky--Kruskal experiment has revealed, amongst other things, the existence of hidden solitons in the wave profile. Here, using Osborne's nonlinear Fourier analysis, which is based on the periodic, inverse scattering transform, the hidden soliton hypothesis is corroborated, and the \emph{exact} number of solitons, their amplitudes and their reference level is computed. Other ``less nonlinear'' oscillation modes, which are \emph{not} solitons, are also found to have nontrivial energy contributions over certain ranges of the dispersion parameter. In addition, the reference level is found to be a non-monotone function of the dispersion parameter. Finally, in the case of large dispersion, we show that the one-term nonlinear Fourier series yields a very accurate approximate solution in terms of Jacobian elliptic functions.
\end{abstract}

\begin{keyword}
Hidden solitons \sep Korteweg--de Vries equation \sep Inverse scattering transform \sep Nonlinear Fourier analysis 
\PACS 05.45.Yv \sep 02.30.Ik
\MSC 35Q51 \sep 35P25 
\end{keyword}

\end{frontmatter}


\section{Introduction}
\label{sec:intro}

Four decades after the discovery of solitons by Zabusky and Kruskal (ZK) \cite{zk65} through a computational experiment, the study of the evolution of harmonic initial data under the Korteweg--de Vries (KdV) equation on a periodic interval is far from complete \cite{cl08}. Beyond the discovery \cite{zk65} of the ability of the KdV equation's (localized) traveling-wave solutions (termed `solitons') to retain their ``identity'' (shape, speed) after collisions, modern numerical simulations by Salupere {\it et al.}\ of this paradigm equation of solitonics reveal such exotic features as ``hidden'' (or ``virtual'') solitons \cite{es05}, emergence of soliton ensembles \cite{spe02} and long-time periodic patterns of the trajectories \cite{sep03,spe03}. These phenomena have been shown to be generic of nonlinear waves, as they occur under other governing equations as well \cite{is09,se04,seii05}

This raises the simple, yet quite fundamental, question: How many solitons emerge from a harmonic input? A successful approach to answering this question is based upon \emph{discrete spectral analysis} \cite{eskm01,sme94,smek96}. The essence of this method is to characterize the solitary waves based on the information inherent in the pseudospectral numerical approximation of the underlying partial differential equation (PDE) \cite{s09}. In general, solitary wave identification is a difficult problem \cite{nch91,zsf96}, especially when multiple wave interactions occur and long time scales are considered. Fortunately, in the case of the ZK experiment, the KdV equation has an advantage over other nonlinear wave equations in that it is \emph{integrable}, i.e., it can be solved exactly (in theory) using the \emph{inverse scattering transform} \cite{as81} on both the infinite line and on a periodic interval.

In this respect, Osborne's \emph{nonlinear Fourier analysis} \cite{oo91} provides a natural framework for applying the periodic, inverse scattering transform (PIST) for the KdV equation to real-world problems. In particular, it overcomes the difficulty of the PIST being only a theoretical tool by providing a practical numerical implementation of it \cite{o91,o94,os93}. Osborne and Bergamasco \cite{ob86} employed this approach to successfully reproduce the numerical results of Zabusky and Kruskal \cite{zk65}. They were able to confirm the number of solitons observed emerging from a harmonic input and the recurrence time of the initial condition. In present work, we employ the same approach to corroborate the numerical results of Salupere {\it et al.}\ regarding hidden solitons. Specifically, the aim here is to determine the hidden modes' amplitudes and classify them in the hierarchy of solutions to the periodic KdV equation (i.e., as either solitons, cnoidal waves or harmonic waves). Conceptually, one may consider this approach as a generalization of some of the ideas in \cite{eskm01,smek96}.

This paper is organized as follows. In Sec.~\ref{sec:posprob}, the physical form of the  Korteweg--de Vries equation and its transformation to the form considered in \cite{zk65} is presented. In Sec.~\ref{sec:pist}, the interpretation of the PIST as nonlinear Fourier analysis is discussed. In Sec.~\ref{sec:numerics}, the PIST spectrum of the harmonic initial condition for the KdV equation is shown for various values of the dispersion parameter, and the hidden soliton hypothesis is discussed. Sec.~\ref{sec:smalldelta}, in the large-dispersion case, illustrates in more detail how a nonlinear Fourier series is constructed using the PIST. Finally, before concluding in Sec.~\ref{sec:concl}, in Sec.~\ref{sec:reflvl} we elaborate on the notion of a soliton reference level for the periodic problem and its dependence on the dispersion parameter.

\section{Position of the problem}\label{sec:posprob}

The KdV equation is arguably the most famous ``soliton-bearing'' PDEs, governing phenomena as seemingly disparate as the motion of lattices, the collective behavior of plasmas and the shape of hydrodynamic waves (see, e.g., \cite{as81,dp06,op94,zk65} and the references therein). The classical context \cite{kdv95} in which it arises is the propagation of ``long'' waves over ``shallow'' water. (For purposes of the present work, we do not need to make these terms any sharper.) Letting $\eta(x,t)$ be the surface elevation, the KdV equation in the moving frame is
\begin{equation}
\eta_t+c_0\eta_x+\alpha\eta\eta_x+\beta\eta_{xxx}=0,\qquad (x,t)\in[0,L]\times(0,\infty),
\label{eq:kdv}
\end{equation}
where $L(>0)$ is the length of the domain, and the subscripts denote partial differentiation with respect to an independent variable. In addition, the speed of linear waves $c_0$, the nonlinearity coefficient $\alpha$ and the dispersion coefficient $\beta$ are constant physical parameters. For {surface water waves}, for example, they can be expressed in terms of the channel depth $h$ and the acceleration due to gravity $g$ as follows  \cite{op94}:
\begin{equation}
c_0 = \sqrt{gh},\qquad \alpha = \frac{3c_0}{2h},\qquad \beta = \frac{c_0h^2}{6}.
\label{eq:kdv_params}
\end{equation}

As in \cite{zk65}, we are interested in the periodic Cauchy problem (i.e., the initial-value problem subject to periodic boundary conditions). Thus, we have that
\begin{equation}
\left\{\begin{aligned}
&\eta(x,0)=\eta_0(x), &&\quad x\in[0,L],\\
&\eta(x+L,t)=\eta(x,t), &&\quad (x,t)\in[0,L]\times[0,\infty).
\end{aligned}\right.
\label{eq:kdv_ic}
\end{equation}
A particularly illustrative initial condition, which is the one we focus on in the present study, is the harmonic one:
\begin{equation}
\eta_0(x) = a \cos(\omega x),\qquad \omega = \frac{2\pi}{L}n,\quad n\in\mathbb{Z},
\label{eq:sin_ic}
\end{equation}
where $a$ is an arbitrary (real) constant.

Now, while Eq.~\eqref{eq:kdv} is in the appropriate form to apply Osborne's nonlinear Fourier analysis, it is not in the form originally considered by Zabusky and Kruskal \cite{zk65} and the subsequent works of Salupere {\it et al.} To make the comparison possible, we introduce the following new variables:
\begin{equation}
\eta(x,t) = \frac{1}{\alpha} u(\xi,\tau),\qquad x = (\xi + c_0t)\!\!\!\mod L,\qquad t=\tau,\qquad \beta=\delta^2, 
\end{equation}
and for definiteness we take $L=2$ cm and $n=1$. Upon substituting the latter transformations into Eq.~\eqref{eq:kdv}, we obtain
\begin{equation}
u_{\tau}+uu_{\xi}+\delta^2u_{\xi\xi\xi}=0,\qquad (\xi,\tau)\in[0,2]\times(0,\infty),
\label{eq:kdv2}
\end{equation}
and the initial--boundary conditions in Eq.~\eqref{eq:kdv_ic} become
\begin{equation}
\left\{\begin{alignedat}{2}
&u(\xi,0)=\alpha a \cos(\pi \xi),&&\qquad \xi\in[0,2],\\
&u(\xi+2,\tau)=u(\xi,\tau),&&\qquad (\xi,\tau)\in[0,2]\times[0,\infty).
\end{alignedat}\right.
\end{equation}
Clearly, we must choose $a = 1/\alpha$ to normalize the initial condition as in \cite{zk65}. 

Also, we must rewrite the parameters in Eq.~\eqref{eq:kdv} in terms of the known quantities $\delta$ and $g=981$ cm/s$^2$. To this end, we use the identity $\beta = \delta^2$ and Eq.~\eqref{eq:kdv_params} to deduce
\begin{equation}
h = \left(\frac{6\delta^2}{\sqrt{g}}\right)^{2/5},\qquad c_0 = \left(6\delta^2g^2\right)^{1/5},\qquad \alpha = \frac{3}{2}\left(\frac{6\delta^2}{g^3}\right)^{-1/5}.
\end{equation}
Finally, we note that the dispersion parameter $\delta$ used here is identical to the one in \cite{zk65}, and it is related to the dispersion parameter $d_l$ of Salupere {\it et al.}\ via $d_l = -2\log_{10}(\pi\delta)$.

\section{Interpretation of the PIST as a ``nonlinear Fourier transform''}\label{sec:pist}

Next, we turn to the relationship between the PIST and the (ordinary) Fourier transform, and the interpretation of the former as a nonlinear generalization of the latter. To this end, first we note that the solution strategy by the scattering transform can be split into two distinct steps: the \emph{direct problem} and the \emph{inverse problem}. The former consists of solving the eigenvalue problem
\begin{equation}
\mathcal{H}\psi = E\psi,\qquad \mathcal{H} := -\frac{\partial^2}{\partial x^2} + V(x),\qquad x\in[0,L],
\label{eq:schro}
\end{equation} 
where $V(x) := -\lambda\eta_0(x)$ is the ``potential,'' $\lambda=\alpha/(6\beta)$ is a nonlinearity-to-dispersion ratio, and $E\in\mathbb{R}$ is a spectral eigenvalue. Equation~\eqref{eq:schro} has been studied extensively: in quantum mechanics it is the celebrated (time-independent) Schr\"odinger equation \cite{as81}, and, in the theory of ODEs, it is known as Hill's equation \cite{c99}. For periodic ``potentials,'' i.e., when $V(x+L)=V(x)$ $\forall x\in[0,L]$ as we have assumed, it is well-known that the spectrum of the operator $\mathcal{H}$ is divided into two distinct sets depending upon the boundary conditions imposed on the eigenfunctions $\psi$ \cite{as81,d05,o94}. Thus, it is common to classify the spectral eigenvalues (also known as the ``scattering data'') as belonging to either the \emph{main spectrum}, which we write as the set $\{\mathcal{E}_j\}_{j=1}^{2N+1}$, or the \emph{auxiliary spectrum}, which we write as the set $\{\mu_j^0\}_{j=1}^{N}$, where $N$ is the number of degrees of freedom (i.e., oscillations modes \cite{oo91} or band gaps \cite{d05}).

On the other hand, the inverse problem consists of constructing the \emph{nonlinear Fourier series} from the spectrum $\{\mathcal{E}_j\}\cup\{\mu_j^0\}$ using either Abelian hyperelliptic functions \cite{o93,os90} or the Riemann $\Theta$-function \cite{bh91,o95}. In former case, which is the so-called $\mu$-representation of the PIST, the \emph{exact} solution of Eq.~\eqref{eq:kdv}, subject to the initial and boundary conditions given in Eq.~\eqref{eq:kdv_ic}, takes the form
\begin{equation}
\eta(x,t)=\frac{1}{\lambda}\left\{2\sum_{j=1}^N \mu_j(x,t) - \sum_{j=1}^{2N+1} \mathcal{E}_j\right\}.
\label{eq:nl_fs}
\end{equation}
It is important to note that all \emph{nonlinear} waves and their \emph{nonlinear} interactions are accounted for in this \emph{linear} superposition. Unfortunately, the computation of the nonlinear oscillation modes (i.e., the hyperelliptic functions $\{\mu_j(x,t)\}_{j=1}^N$) is \emph{highly} nontrivial; however, numerical approaches have been developed \cite{os90} and successfully used in practice \cite{op94,osbc91}. 

Several special cases of Eq.~\eqref{eq:nl_fs} offer insight into why the latter is analogous to the ordinary Fourier series and aid with the interpretation of the results in the following sections. In the small-amplitude limit, i.e., when $\max_{x,t}|\mu_j(x,t)|\ll1$, we have $\mu_j(x,t)\sim \cos(k_j x-\omega_j t+\phi_j)$, where $k_j$ is the wavenumber, $\omega_j$ is the frequency and $\phi_j$ the phase of the mode. Therefore, if we suppose that all the oscillation modes fall in the small-amplitude limit, then Eq.~\eqref{eq:nl_fs} reduces to the ordinary Fourier series! This relationship is more than just an analogy, a rigorous derivation of the (ordinary) Fourier transform from the scattering transform, in the small-amplitude limit, is given in \cite{o91}. Next, if there are no interactions, e.g., the spectrum consists of a single wave (i.e., $N=1$), we have $\mu_1(x,t) = \cn^2(k_1 x-\omega_1 t+\phi_1|m_1),$ which is a Jacobian elliptic function with modulus $m_1$. In fact, it is the well-known \emph{cnoidal wave} solution of the (periodic) KdV equation \cite{as81}.

For the hyperelliptic representation of the nonlinear Fourier series, given by Eq.~\eqref{eq:nl_fs}, the wavenumbers are \emph{commensurable} with those of the ordinary Fourier series, i.e., $k_j=2\pi j/L$ ($1\le j\le N$) \cite{o93}. However, this is not the only way to classify the nonlinear oscillations. One can use the modulus $m_j$, termed the ``soliton index,'' of each of the hyperelliptic functions, which can be computed from the main spectrum as 
\begin{equation}
m_j=\frac{\mathcal{E}_{2j+1}-\mathcal{E}_{2j}}{\mathcal{E}_{2j+1}-\mathcal{E}_{2j-1}},\qquad 1\le j\le N.
\label{eq:modulus}
\end{equation}
Then, each nonlinear oscillation mode can be placed into one of two distinct categories based on its soliton index:
\begin{enumerate}
	\item $m_j\gtrsim0.99$ $\Rightarrow$ solitons, in particular, $\cn^2(\zeta|m=1)=\sech^2(\zeta)$;
	\item $m_j\ll1.0 \Rightarrow$ radiation, in particular, $\cn^2(\zeta|m=0)=\cos^2(\zeta)$.
\end{enumerate}
For $m_j$ not in either distinguished limit above, the qualitative structure of the nonlinear oscillations is not immediately obvious. Boyd \cite{b82} argues that for moderate moduli the polycnoidal wave solutions of the KdV equation are actually well-approximated by the first few terms in their Fourier series, showing they are, in fact, not much different from the linear waves of the $m_j\ll 1.0$ limit.

Furthermore, it can be shown \cite{op94,osbc91} that the amplitudes of the hyperelliptic functions are given by
\begin{equation}
A_j = \left\{\begin{alignedat}{2} &\frac{2}{\lambda}(E_{\mathrm{ref}}-\mathcal{E}_{2j}),&&\quad \text{for solitons};\\ &\frac{1}{2\lambda}(\mathcal{E}_{2j+1}-\mathcal{E}_{2j}), &&\quad \text{otherwise (radiation)}.\end{alignedat}\right.
\label{eq:Aj}
\end{equation}
where $E_{\mathrm{ref}}=\mathcal{E}_{2j^*+1}$ is the \emph{soliton reference level} with $j^*$ being the largest $j$ for which $m_j\ge 0.99$. Then, clearly, the number of solitons in the spectrum is $N_{\mathrm{sol}}\equiv j^*$.

Finally, we note that the theory of Eq.~\eqref{eq:schro} is quite mature and exact solutions can be obtained for a number of specific forms of the potential $V(x)$ \cite{as81,c99}. Unfortunately, for the ZK experiment (recall Eq.~\eqref{eq:sin_ic}), we have $V(x)=-\lambda a \cos(\omega x)$  for which there is no known closed form solution. All is not lost, however, the essence of Osborne's nonlinear Fourier analysis is that a numerical solution of Eq.~\eqref{eq:schro} can be obtained and the details of the PIST carried out in this way. To this end, we use a modified version of Osborne's automatic algorithm \cite{o94}, as described in \cite{c09}. The result is an exact (to any desired numerical precision) representation of the nonlinear Fourier spectrum of any initial condition of the KdV equation.

\section{How many solitons in a cosine wave?}\label{sec:numerics}
\subsection{The Zabusky--Kruskal experiment}

\begin{figure}
\centerline{\includegraphics[width=0.475\textwidth]{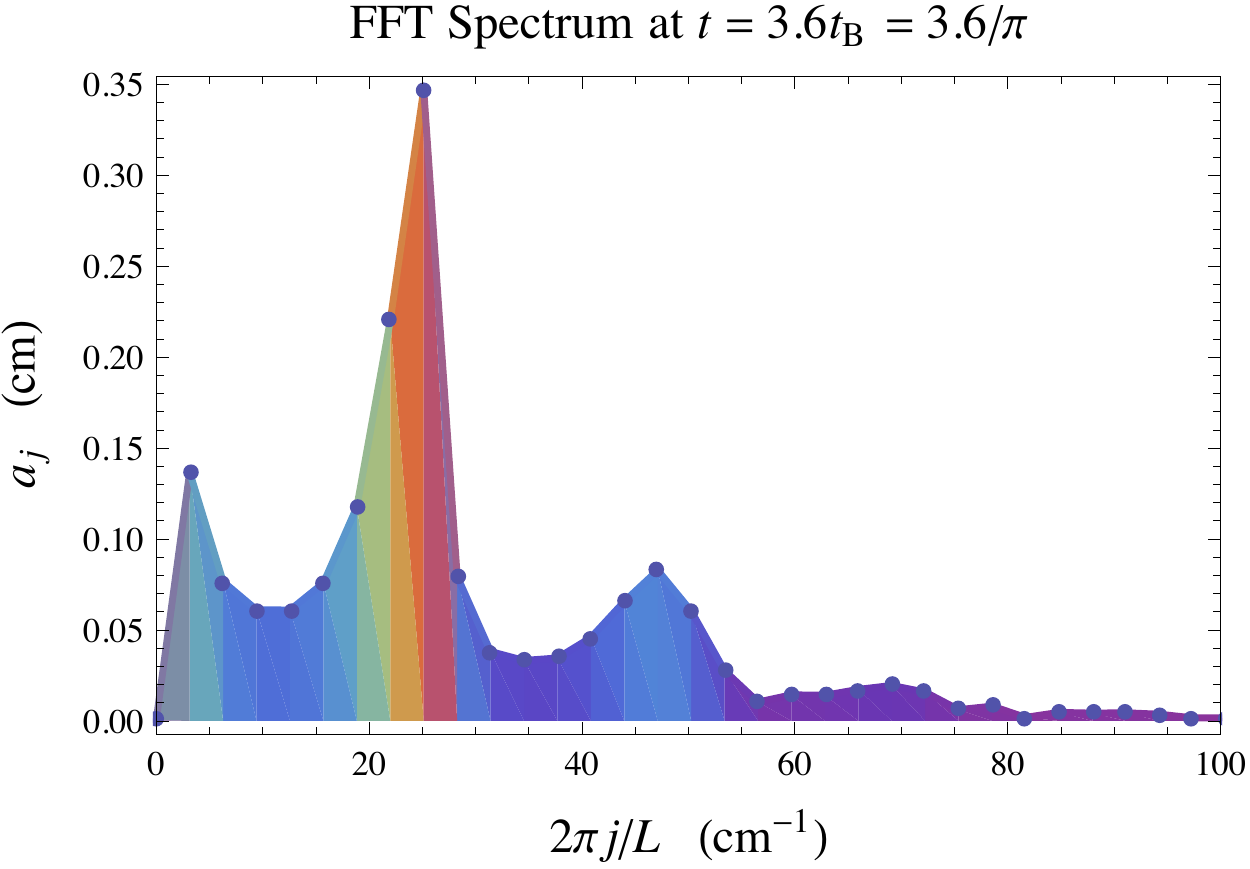}\hspace{8mm}\includegraphics[width=0.475\textwidth]{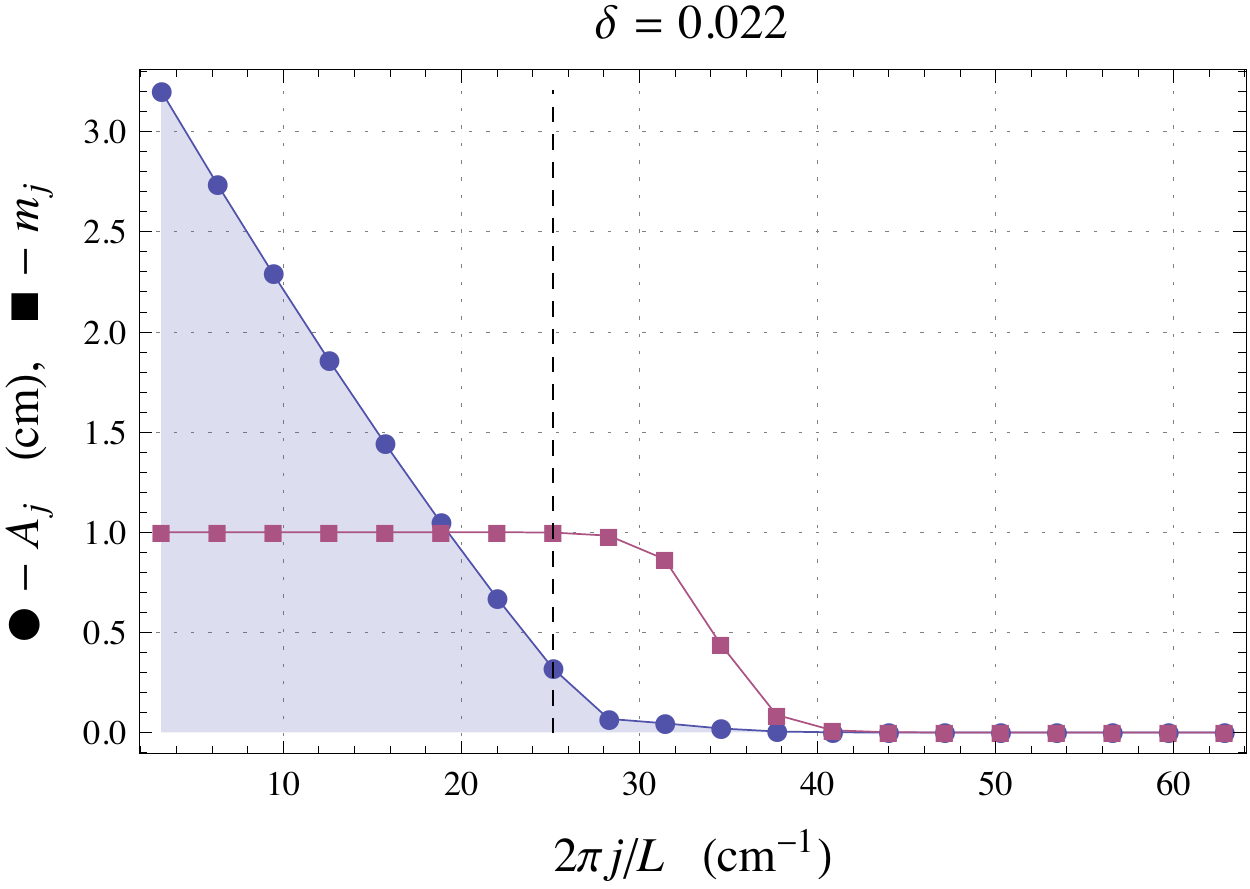}}
\caption{(Color online.) Comparison of the ordinary Fourier spectrum (left panel) at a specific time and the nonlinear Fourier spectrum (right panel) of the original ZK experiment ($\delta = 0.022$). The vertical dashed line in the right panel denotes the wave number of the last true soliton mode.}
\label{fig:zk_spectrum}
\end{figure}

Recall that, in Sec.~\ref{sec:posprob} upon switching to the new set of variables, we distilled all physical parameters of the KdV equation into the dispersion parameter $\delta$. Therefore, it suffices to vary $\delta$ to establish all possible ways a harmonic initial condition can evolve under the KdV equation. In this subsection, we analyze the classical ZK experiment, i.e., $\delta = 0.022$, using the PIST. To this end, in Fig.~\ref{fig:zk_spectrum}, the ordinary and nonlinear Fourier spectra of the harmonic initial condition are presented. Since the ordinary Fourier transform, implemented here using the Fast Fourier Transform (FFT), does not resolve the temporal evolution of a solution of the KdV equation, the FFT of the solution, which was computed by direct numerical integration of the PDE, is presented at $t = 3.6/\pi$. This temporal value was chosen for ease of comparison with \cite{zk65} and because all observable solitons are visible in the wave profile at this instant of time.

The FFT spectrum suggests that there are \emph{over thirty} significant normal modes of the problem, with the majority of the enegry concentrated in three distinct wave number bands centered at $k\approx2.5$ cm$^{-1}$, $k\approx25$ cm$^{-1}$ and $k\approx47.5$ cm$^{-1}$. Of course, without further study of the FFT spectrum and its evolution in time, nothing can be said about the number of solitons present in the solution profile.

In contrast, the PIST shows \emph{exactly eight} soliton modes plus other nonlinear waves and radiation. As was shown in \cite{ob86}, the PIST spectrum captures precisely the eight solitons observed in \cite{zk65}. However, what has never been discussed before are the other (four of them, in fact, as Table~\ref{tb:sols} below shows) nontrivial modes in the spectrum. By `nontrivial' we mean their amplitudes are not so small as to have negligible energy contribution. The discrete spectral analysis approach of Salupere {\it et al.}\ \cite{sme94,smek96} found there exist at least four hidden ``solitons'' beyond the eight observable ones. Indeed, in Fig.~\ref{fig:zk_spectrum}, an additional four modes beyond the ``true'' soliton ones are easily distinguished (i.e., have large enough amplitude). Thus, the hidden ``soliton'' hypothesis is true. What is more, the PIST naturally classifies these hidden \emph{modes} within the hierarchy of solutions to the periodic KdV equation; it happens they are \emph{not} proper solitons (i.e., they have $m_j < 0.99$).

Finally, Fig.~\ref{fig:zk_spectrum} illustrates a very important point: namely, that the FFT is only useful when analyzing a \emph{solution}. That is to say, the underlying PDE has to be solved up to some instant of time, and then the spectrum computed from this data. Other tools, that can provide both time and frequency resolution, such as the \emph{short-time} (or \emph{windowed}) Fourier transform exist \cite{g01}, however, the PDE must still be solved in advance. On the other hand, the PIST fully characterizes the \emph{initial condition} and its \emph{evolution}. This is due to the fact that the PIST is a formal technique for integrating exactly the KdV and other such integrable nonlinear wave equations.

\subsection{The effect of dispersion on soliton generation}

\begin{figure}[!ht]
\centerline{\includegraphics[width=0.475\textwidth]{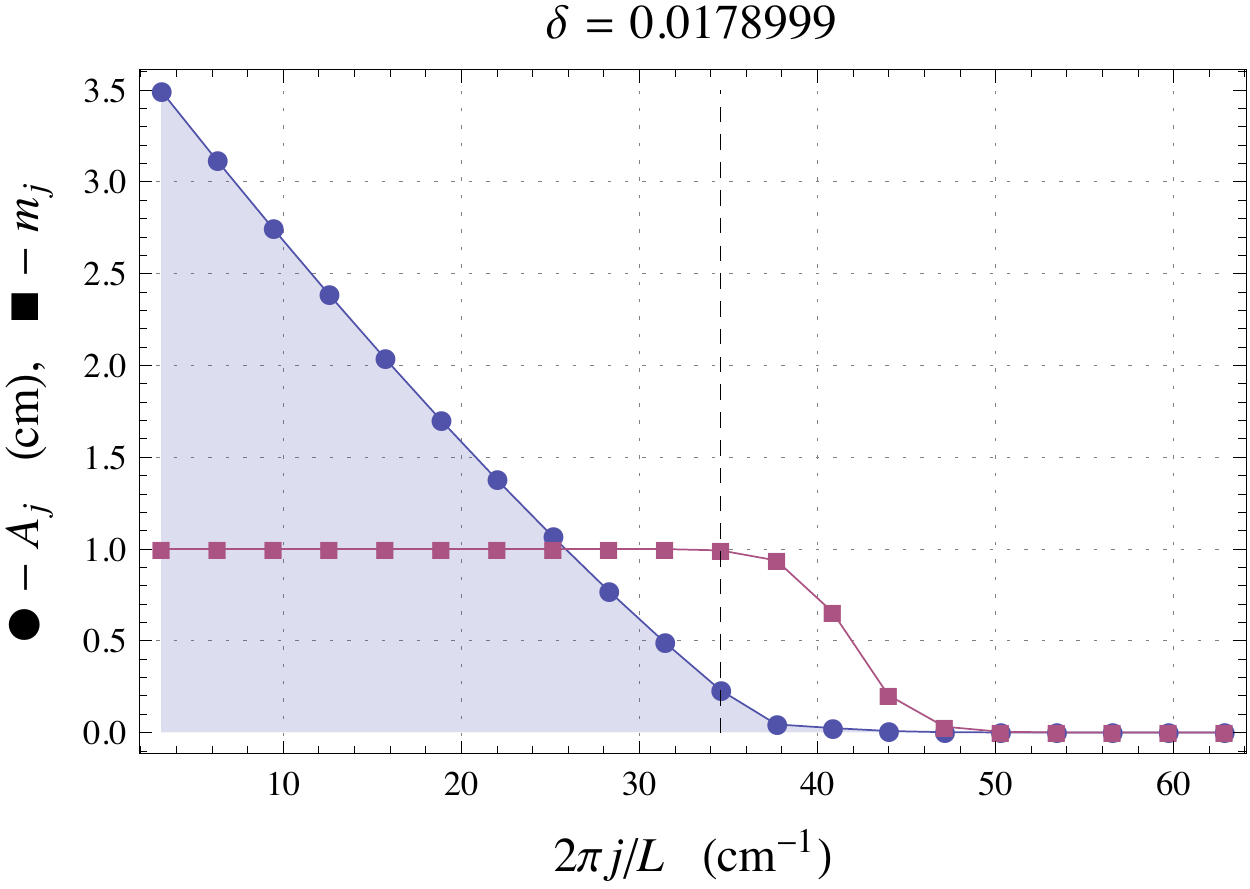}\hspace{8mm}\includegraphics[width=0.475\textwidth]{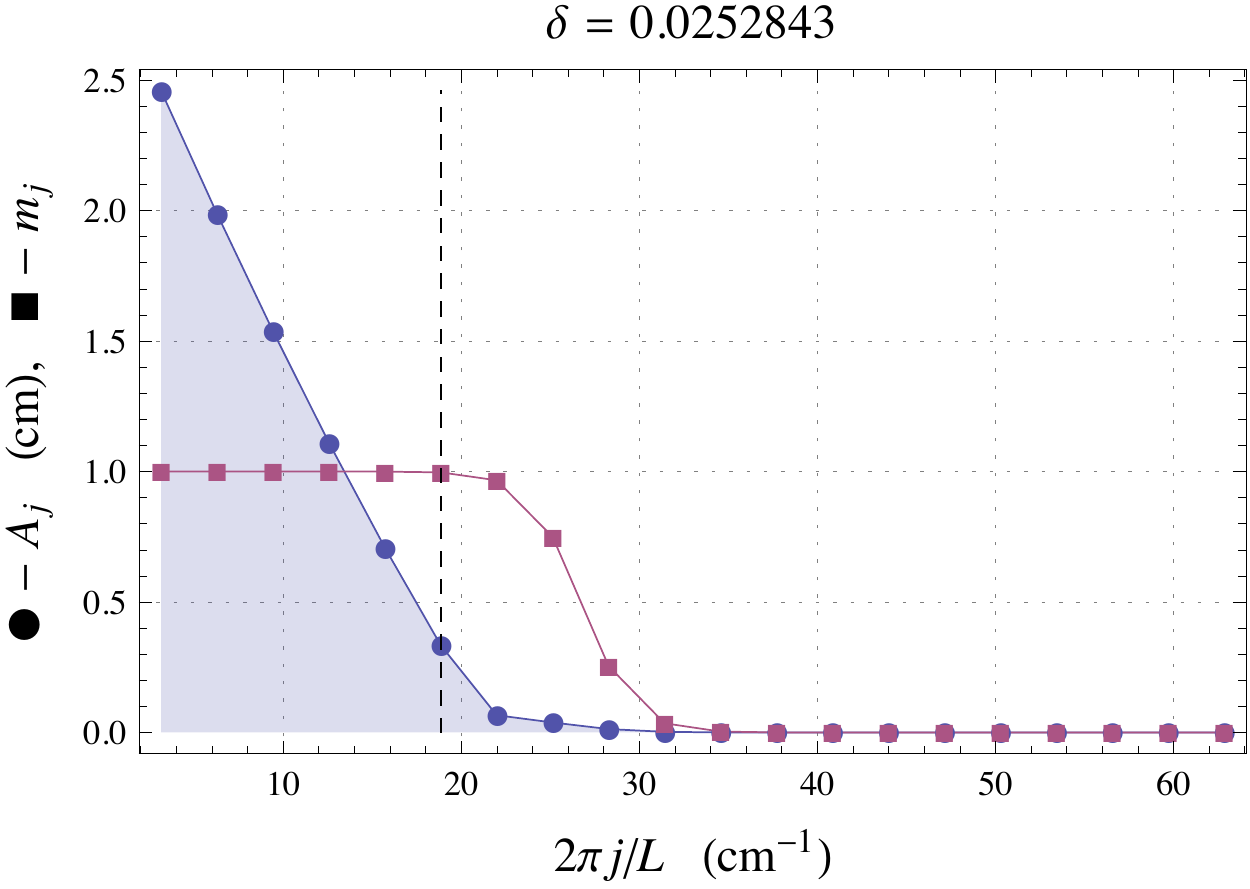}}
\vspace{5mm}
\centerline{\includegraphics[width=0.475\textwidth]{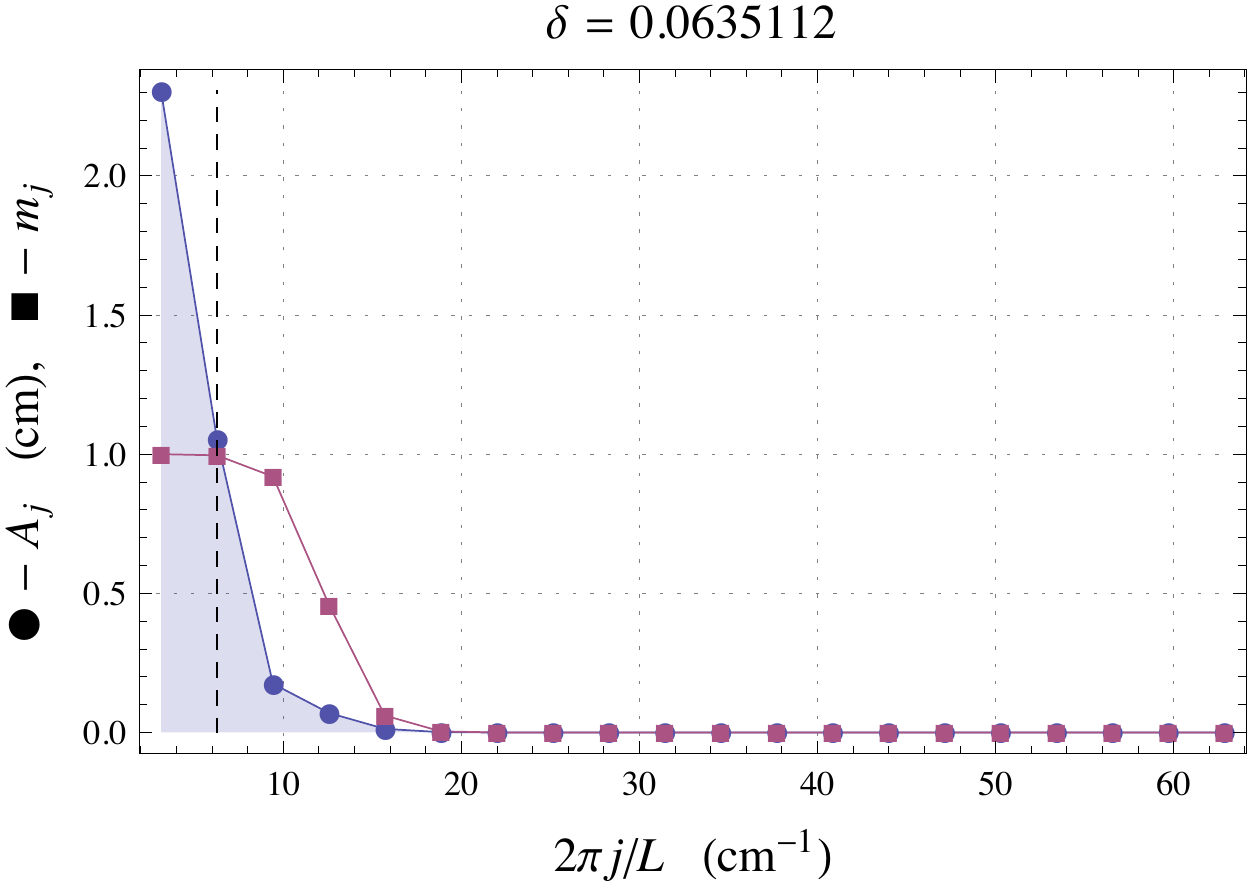}\hspace{8mm}\includegraphics[width=0.475\textwidth]{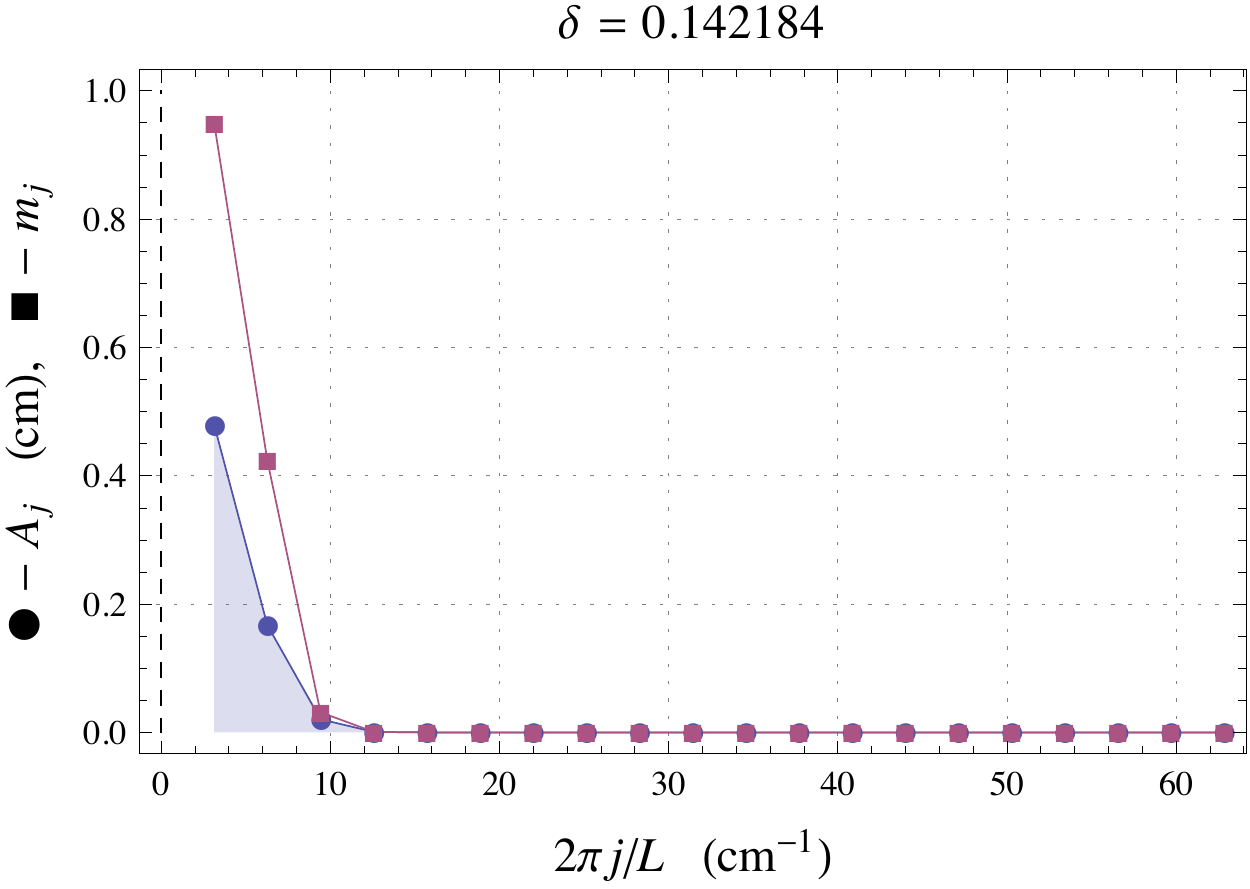}}
\caption{(Color online.) Nonlinear Fourier spectra of the harmonic initial condition for four representative values of $\delta$. As before, a vertical dashed line delineates the last soliton mode from the remainder of the spectrum.}
\label{fig:moderate_delta}
\end{figure}

It has been shown \cite{eskm01,sme94,smek96} that the number of solitons detected in the solution of the KdV equation varies with the dispersion parameter. Indeed, it is well-known that, as the zero-dispersion limit ($\delta\to0$) is approached, the number of solitons generated by an initial condition grows quickly \cite{gk07}. In Fig.~\ref{fig:moderate_delta}, we present the PIST spectra of the solution for a number of representative values of $\delta$ besides the ZK value; these were chosen so that a comparison with \cite{eskm01,sme94,smek96} is possible. The quantitative results from theses figures, and all other ones presented in the his paper, are also summarized in Table~\ref{tb:sols}, which includes the relevant data from \cite{eskm01,sme94,smek96} and the heuristic estimate $N_{\mathrm{sol}} \approx 0.2/\delta$ due to Zabusky \cite{z81}.

For $\delta = 0.0178999$ ($\Leftrightarrow d_l=2.5$), we observe eleven solitons and nine non-soliton nonlinear waves for a total of twenty waves generated by the harmonic initial condition. Though this value of $\delta$ was too small for the computational efforts at the time of the studies in \cite{eskm01,smek96}, the PIST has no issues whatsoever with $\delta\to0$. The computational time for the nonlinear Fourier analysis grows only linearly with the number of spectral eigenvalues, and it is independent of $\delta$.

For $\delta = 0.0252843$ ($\Leftrightarrow d_l=2.2$), we observe six solitons and eight non-soliton nonlinear waves. For $\delta = 0.0635112$ ($\Leftrightarrow d_l=1.4$), the PIST finds two solitons and six other modes. And, for $\delta = 0.142184$ ($\Leftrightarrow d_l=0.7$), there are zero solitons out of a total of five waves in the spectrum. The pattern is clear: as the dispersion parameter increases the effects of nonlinearly weaken and fewer solitons are produced from the initial harmonic wave. The number of other non-soliton modes in the spectrum also decreases with $\delta$, from nine at $\delta= 0.0178999$ to five at $\delta = 0.142184$.

Now, returning to Table~\ref{tb:sols} and comparing the number of solitons $N_{\mathrm{sol}}$ given by the PIST and the number of \emph{visible} solitons observed by Salupere {\it et al.}\ (i.e., total minus hidden), we see that \emph{all but one} of the observed ``solitons'' in the wave profile are indeed solitons, and the last is a highly-nonlinear cnoidal wave. This shows that across a wide range of $\delta$ values, as was the case for the ZK experiment discussed in the previous subsection, the hidden ``solitons'' are a manifestation of the other nonlinearly interacting non-soliton solutions of the periodic KdV equation.
\begin{table}[!ht]
\begin{center}
\begin{tabular}{llcccccccc}
\toprule
& & & \multicolumn{2}{c}{Salupere {\it et al.}} & & & &\\
\cmidrule(r){4-5}
$\delta$ & $d_l$ & $\round(0.2/\delta)$ & Total & Hidden & $N$ & $N_{\mathrm{sol}}$ & $N_{A_j > 10^{-3}}$ & $N_{m_j > 10^{-1}}$\\
\midrule
$0.0178999$ & $2.5$ & 11 & -- & -- & 20 & 11 & 15 & 14\\
\hline
$0.022$ & $2.320854$ & 9 & 12 & 3 & 17 & 8 & 12 & 11\\
\hline
$0.0252843$ & $2.2$ & 8 & 10 & 3 & 14 & 6 & 10 & 9\\
\hline
$0.0635112$ & $1.4$ & 3 & 4 & 1 & 8 & 2 & 6 & 4\\
\hline
$0.142184$ & $0.7$ & 1 & 2 & 1 & 5 & 0 & 3 & 2\\
\hline
$0.317998$ & $-8.5\times10^{-4}$ & 1 & -- & -- & 3 & 0 & 2 & 1\\
\hline
$1.0$ & $-0.9943299$ & 0 & -- & -- & 3 & 0 & 2 & 0\\
\bottomrule
\end{tabular}
\end{center}
\caption{Comparison of the discrete spectral and nonlinear Fourier analyses of hidden ``solitons'' and extended statistics of the nonlinear Fourier spectrum for all values of $\delta$ considered here.}
\label{tb:sols}
\end{table}

Finally, we note that one cannot expect to detect (in, e.g., a numerical experiment) \emph{all} nonlinear oscillation modes the PIST finds. That is, the discrete spectral analysis \cite{eskm01,smek96} is inherently limited in its ability to resolve low amplitude, or very weakly nonlinear waves in the spectrum. Therefore, to shed some light on precisely \emph{which} modes can be detected and \emph{why}, the last two column of Table~\ref{tb:sols} show the number $N_{A_j > 10^{-3}}$ of modes with amplitudes greater than $10^{-3}$ and the number $N_{m_j > 10^{-1}}$ of modes with elliptic moduli greater than $10^{-1}$, respectively. Though these cutoffs are largely arbitrary, they are qualitatively reasonable. Indeed, the \emph{total} number of ``solitons'' found by Salupere {\it et al.}\ is either of these numbers, for the values of $\delta$ considered here. Thus, the ``small'' and ``weak'' modes in the spectrum (typically, these are radiation modes) are difficult to resolve through discrete spectral analysis. However, we have shown, beyond any doubt, that the discrete spectral analysis properly distinguishes the total number of \emph{highly} nonlinear oscillation modes, both soliton and otherwise, and that there indeed are hidden modes in the ZK experiment.

\section{A note on the large dispersion case}\label{sec:smalldelta}

\begin{figure}[!ht]
\centerline{\includegraphics[width=0.475\textwidth]{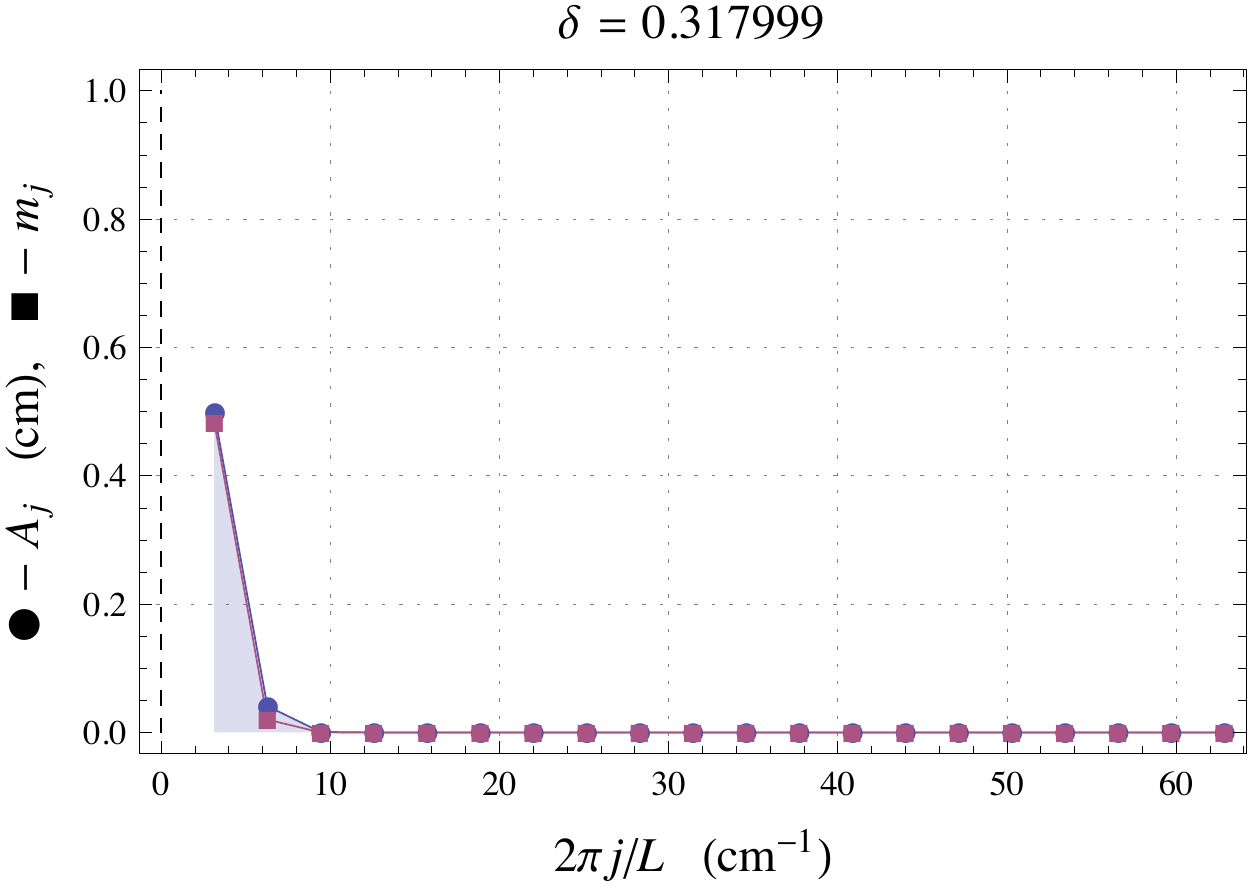}\hspace{8mm}\includegraphics[width=0.475\textwidth]{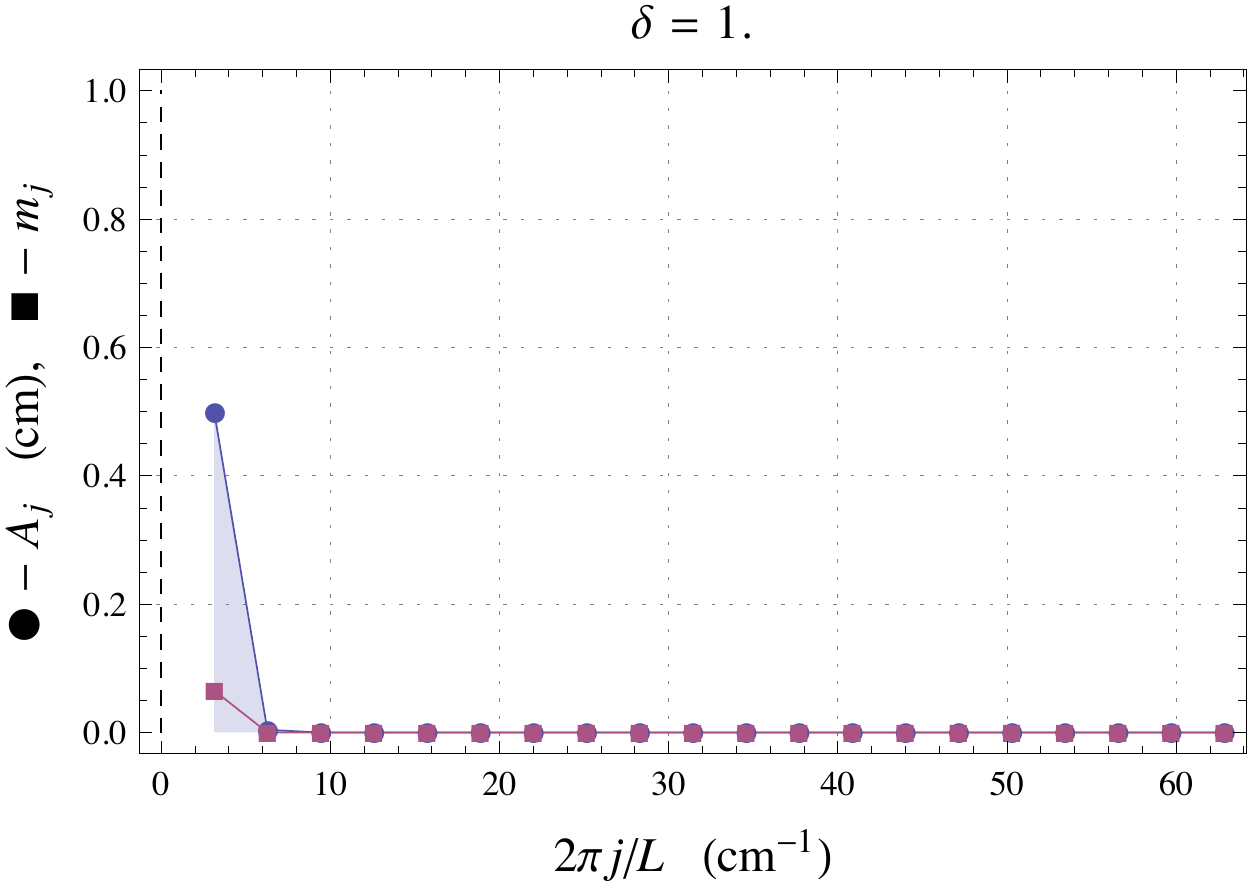}}
\caption{(Color online.) Nonlinear Fourier spectra of the harmonic initial condition for large values of the dispersion parameter. The vertical dashed line representing the end of the soliton spectrum is now at $k=0$ because solitons do not emerge in the solution for these values of $\delta$.}
\label{fig:large_delta}
\end{figure}

Though, from a mathematical point of view, the zero-dispersion ($\delta\to0$) limit is by far the most interesting distinguished limit of the KdV equation \cite{gk07}, the large-dispersion one ($\delta=\mathcal{O}(1)$) illustrates very well some of the concepts behind the PIST and Osborne's nonlinear Fourier analysis. To this end, Fig.~\ref{fig:large_delta} shows the nonlinear Fourier spectrum of the harmonic initial conditions for two large values of $\delta$. Clearly, as $\delta\to1$, there is only one large-amplitude mode, with the remaining two modes' amplitudes approaching zero as $\delta$ increases (recall Table~\ref{tb:sols}). Moreover, the ``degree of nonlinearity'' (i.e., the elliptic modulus $m_j$) of the leading mode decreases as $\delta$ increases, while its amplitude remains the same. This is due to the fact that the terms $\delta^2u_{\xi\xi\xi}$ and $uu_\xi$ in Eq.~\eqref{eq:kdv2} now balance asymptotically, and there is no nonlinear steepening of the wave profile that leads to the generation of solitons \cite{zk65}.

\begin{figure}[!ht]
\centerline{\includegraphics[width=0.475\textwidth]{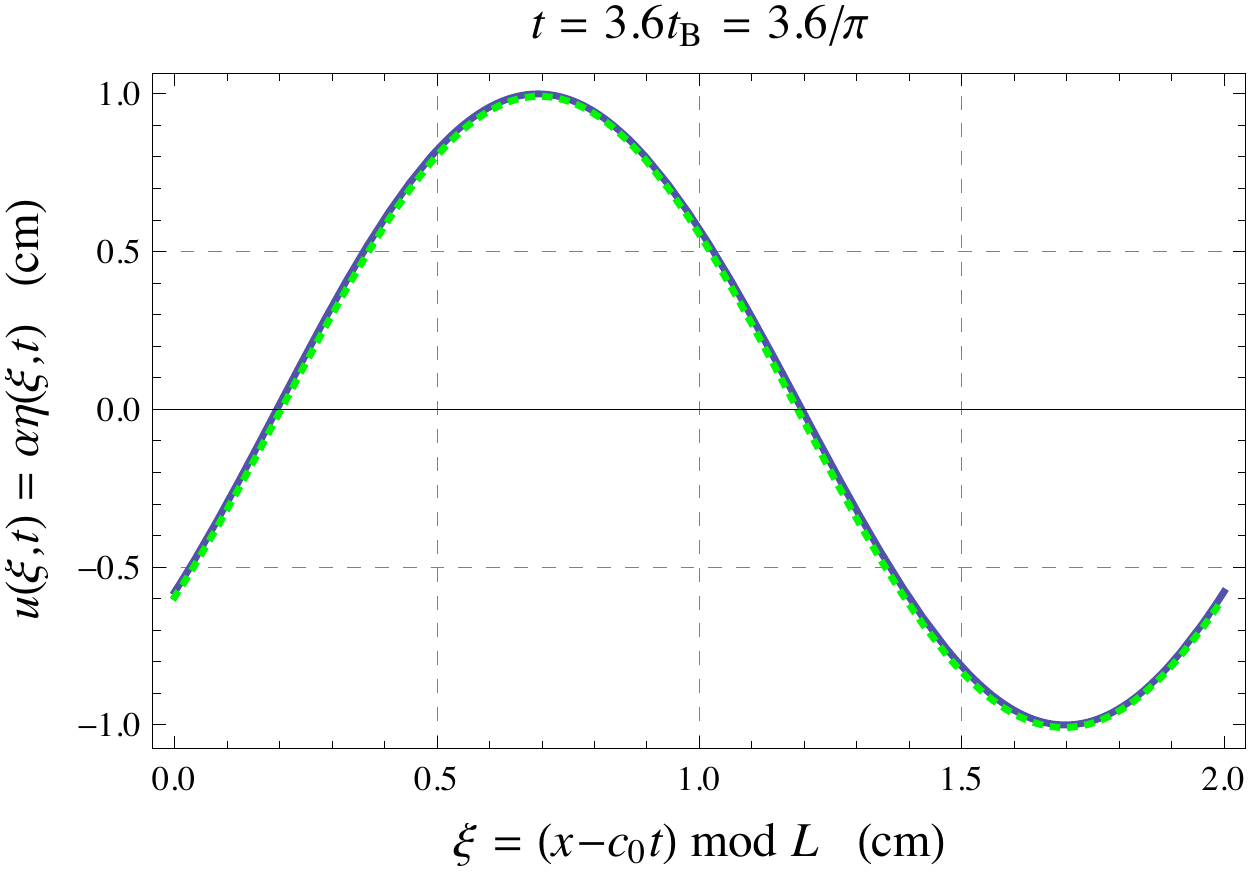}}
\caption{(Color online.) Comparison between the numerical (blue) and approximate PIST solution (dashed green), given by Eq.~\eqref{eq:1mode}, of the large-dispersion KdV equation ($\delta=1$).}
\label{fig:1mode}
\end{figure}

Thus, to a good approximation, $N=1$ when $\delta=\mathcal{O}(1)$. In this case, the solution of the periodic KdV equation takes the following form \cite{o93}:
\begin{equation}
u(\xi,\tau) \approx 4A_1\cn^2(k_1\xi+\omega_1 \tau|m_1) +\lambda^{-1}(\tilde{\mathcal{E}} - 2\mathcal{E}_3),
\label{eq:1mode}
\end{equation}
where $k_1 = \sqrt{\mathcal{E}_3-\mathcal{E}_1}$ is the mode's wavenumber, $\omega_1 = 2\tilde{\mathcal{E}}k_1$ is its frequency, and we have set $\tilde{\mathcal{E}} := \sum_{j=1}^3 \mathcal{E}_j$ for convenience. Of course, $\{\mathcal{E}_j\}_{j=1}^3$, $A_1=0.499987$ and $m_1=0.0653120$ are obtained from the PIST. This approximate solution is compared to the numerical solution of the KdV equation, for $\delta=1$, in Fig.~\ref{fig:1mode}. The agreement between the two is quite good. More importantly, Eq.~\eqref{eq:1mode} illustrates a fundamental difference between the \emph{infinite-line} and \emph{periodic} versions of the IST. Not only does the periodic problem have a much richer solution space (a continuum of solutions expressible as Jacobian elliptic functions), but it also requires \emph{three} spectral eigenvalues to determine \emph{one} mode. This means that the infinite-line IST approach in \cite{eskm01,smek96}, while helpful, cannot capture the full picture of the periodic problem.

\section{A reference level for the periodic problem}\label{sec:reflvl}

Recall that in Eq.~\eqref{eq:Aj}, we defined the quantity $E_{\mathrm{ref}}$ as the \emph{soliton reference level}. This idea was first formulated in \cite{ob86}, where Osborne and Bergamasco showed that a shift in the reference (or zero) level of a multi-soliton solution is mathematically equivalent to a shift in the energy level of the last soliton's band gap edge with respect to $E=0$. This gives the natural definition $E_{\mathrm{ref}} = \mathcal{E}_{2j^*+1} - 0$, where (as in Sec.~\ref{sec:pist}) $j^*$ is the largest $j$ for which $m_j\ge 0.99$. Hence, if we know $E_{\mathrm{ref}}$, then we can determine the solitons' reference propagation level and reference wavenumber for the periodic problem as follow:
\begin{equation} 
u_{\mathrm{ref}} = - \alpha E_{\mathrm{ref}}/\lambda,\qquad k_{\mathrm{ref}} = 2\pi j^*/L. 
\label{eq:reflvl}
\end{equation}

Table~\ref{tb:uref} gives $u_{\mathrm{ref}}$ as a function of the dispersion parameter for all values of $\delta$ discussed in the preceding sections. Naturally, in the large-dispersion case no solitons form, and the harmonic initial condition's shape is largely preserved as it propagates. In this limit, the propagation level is most naturally understood as the \emph{mean level} of the solution. Indeed, it is very close to zero---the mean of a trigonometric function over its period. This leads to an important observation: in the presence of large radiation modes in the spectrum, the reference level is \emph{not} simply the minimum of the wave profile in the solution. That is to say, the true $u_{\mathrm{ref}}$ \emph{cannot} be observed in the numerical solution of the problem, as the oscillatory radiation modes will continually shift the wave profile minimum; if they could somehow be ``filtered'' one would see the ``pure'' solitons propagating on the level given by $u_{\mathrm{ref}}$ in Table~\ref{tb:uref}.
\begin{table}[!ht]
\begin{center}
\begin{tabular}{|c||l|l|l|l|l|l|l|}
\hline
$\delta$\rule{0pt}{10pt} & $1.0$ & $0.318$ & $0.142$ & $\phantom{-}0.0635$ & $\phantom{-}0.0253$ & $\phantom{-}0.022$ & $\phantom{-}0.0179$\\
\hline
$u_{\mathrm{ref}}$\rule{0pt}{10pt} & $1.29\times10^{-4}$ & $7.80\times10^{-4}$ & $0.337$ & $-0.483$ & $-0.555$ & $-0.720$ & $-0.844$\\
\hline
\end{tabular}
\end{center}
\caption{The soliton reference level $u_{\mathrm{ref}}$, computed from the PIST spectrum using Eq.~\eqref{eq:reflvl}, for various values of the dispersion parameter $\delta$.}
\label{tb:uref}
\end{table}

What is more interesting, however, is that $u_{\mathrm{ref}}$ increases with $\delta$, until at a certain value it reverses itself and goes back through zero becoming negative. As $\delta \to 0$, it appears that the PIST predicts that $u_{\mathrm{ref}}\to-1$, which is the ``absolute minimum'' level observed numerically \cite{eskm01,sme94,smek96}. This shows that the PIST and direct numerical simulation approaches are consistent, as they should, but that the PIST offers a deeper insight into how $u_{\mathrm{ref}}$ varies with $\delta$, which is quite difficult to infer from the numerical simulation.

Finally, we note that the \emph{infinite-line} inverse scattering transform (IST) was used in \cite{eskm01,smek96} to corroborate the discrete spectral analysis (numerical) approach. But, because the periodic problem for the KdV equation is not mathematically equivalent to the infinite-line problem with an initial condition of compact support, the choice of $u_{\mathrm{ref}}$ in \cite{eskm01,smek96} had to be made \emph{before} the number of solitons could be calculated. This lead to an artificial dependence of $N_{\mathrm{sol}}$ on $u_{\mathrm{ref}}$. Though good results were obtained using an appropriate estimate for $u_{\mathrm{ref}}$, it is clear that the PIST approach, which is the natural one for the periodic problem, gives an unambiguous way to calculate $u_{\mathrm{ref}}$.

\section{Conclusion}\label{sec:concl}

The present work shows that the evolution (``spectrum'') of harmonic initial data under the Korteweg--de Vries (KdV) equation can be fully and automatically characterized by the periodic, inverse scattering transform (PIST). In particular, within the framework of Osborne and Bergamasco \cite{ob86}, the hidden soliton hypothesis of Salupere {\it et al.}\ is corroborated, and the \emph{exact} number of solitons, their amplitudes and their reference level is computed. This offers new insight into the phenomenon because such precise results were not possible through the direct numerical integration of the KdV equation and the discrete spectral analysis approach \cite{es05,eskm01,s09,sep03,sme94,smek96,spe02,spe03}. In particular, the apparent linear variation of the soliton amplitudes with the wavenumber, first discussed in \cite{zk65}, is true for all soliton modes detected by the PIST, over a range of $\delta$ values. Meanwhile, the remainder of the spectrum consists of other ``less nonlinear'' oscillation modes, e.g., nonlinearly-interacting cnoidal waves. Though the amplitudes of these do not follow the linear trend, they are not negligible, and they are, in fact, what is observed numerically as ``hidden solitons.''

It is important to note that the present approach (via the PIST and Osborne's nonlinear Fourier analysis) to the problem of soliton formation is complementary to the discrete spectral analysis approach of Salupere {\it et al.} While the PIST allows for a very precise decomposition of the initial condition into the ``basis'' elements of the nonlinear PDE at hand (in the present case, the KdV equation), it requires that the PDE be fully integrable---a rather stringent stipulation. The discrete spectral analysis approach to soliton formation does not require any {\it a priori} mathematical structure of the PDE.

Finally, two of the findings of the present work require further investigation. First, the non-monotone dependence of $u_{\mathrm{ref}}$ on $\delta$ is quite unexpected. A detailed study is necessary to identify the mechanism for this. Second, the number $N_{\mathrm{sol}}$ of solitons predicted by the PIST is \emph{always} less than the number of ``solitons'' found in the numerical solution by Salupere {\it et al.}\ across the entire range of values of $\delta$ considered here. This means that most hidden ``solitons'' are not solitons, {\it per se}. Clearly, the richness of the solution space of the periodic KdV equation allows for modes that fall ``in-between'' solitons and radiation. Whether this distinction can be made using the discrete spectral analysis is a very interesting question that would be very relevant when studying \emph{non-integrable} wave equations, for which a nonlinear Fourier transform cannot be constructed, and therefore the distinction between soliton and non-soliton modes cannot be made analytically.

\section*{Acknowledgments}

The author is indebted to Prof.~Andrus Salupere for introducing him to this problem and many stimulating discussions on the subject. 
A Student Award from Prof.~Thiab Taha, organizer of the Sixth IMACS International Conference on Nonlinear Evolution Equations and Wave Phenomena: Computation and Theory, which allowed the author to present this work at the latter conference, is kindly acknowledged.
This work was also supported, in part, by a Walter P. Murphy Fellowship from the Robert R.\ McCormick School of Engineering and Applied Science at Northwestern University.

\bibliographystyle{elsart-num-sort}
\bibliography{hidden_solitons}

\end{document}